\documentstyle[12pt]{article}
\newcommand{\drm}{\rm d}
\newcommand{\Real}{{\rm Re}}
\newcommand{\Imag}{{\rm Im}}
\newcommand{\scrE}{\cal E}
\newcommand{\scriptp}{\cal P}

\newcommand{\Le}{{\cal L}}

\newcommand{\pp}{/\!\!\!p}

\newcommand{\om}{\omega}

\newcommand{\bb}{\begin{equation}}
\newcommand{\ee}{\end{equation}}
\newcommand{\bega}{\begin{eqnarray}}
\newcommand{\ega}{\end{eqnarray}}
\newcommand{\begae}{\begin{eqnarray*}}
\newcommand{\egae}{\end{eqnarray*}}

\newcommand{\ga}{\gamma}

\newcommand{\sig}{\sigma}
\newcommand{\Sig}{\Sigma}

\newcommand{\rig}{\rightarrow}

\newcommand{\Lef}{\Leftrightarrow}

\newcommand{\al}{\alpha}

\newcommand{\C}{I\!\!\!C}
\newcommand{\La}{\Lambda}
\newcommand{\la}{\lambda}

\newcommand{\Om}{\Omega}

\newcommand{\h}{\hspace*{0.5 cm}}

\newcommand{\dis}{\displaystyle}

\newcommand{\ov}{\overline}

\newcommand{\pa}{\partial}

\newcommand{\cent}{\centerline}
\newcommand{\vs}{\vspace*}

\newcommand{\Vbf}{\mbox{\boldmath $V$}}
\newcommand{\Abf}{\mbox{\boldmath $A$}}
\newcommand{\sbf}{\mbox{\boldmath $s$}}

\newcommand{\vbf}{\mbox{\boldmath $v$}}
\newcommand{\abf}{\mbox{\boldmath $a$}}
\newcommand{\wbf}{\mbox{\boldmath $w$}}
\newcommand{\xbf}{\mbox{\boldmath $x$}}
\newcommand{\ubf}{\mbox{\boldmath $u$}}
\newcommand{\rbf}{\mbox{\boldmath $r$}}
\newcommand{\impulse}{\mbox{\boldmath $p$}}

\begin{document}
 
\baselineskip 0.6cm
 
\cent{\large\bf SPIN AND ZITTERBEWEGUNG}
 
\cent{\large\bf IN A FIELD THEORY OF THE ELECTRON.$^{(*)}$}

\footnotetext{$^{(*)}$ Work supported by CAPES (Brazil), and by INFN 
and MIUR (Italy)}
 
\vs{0.3cm}
 
\begin{center}
 
{{\bf Erasmo RECAMI} $^{(**)}$

\footnotetext{$^{(**)}$ {\em Bolsista CAPES/BRASIL:} Supported as a Vising Professor by a PVE Fellowship of CAPES/BRAZIL.}

{\em Facolt\`a di Ingegneria, Universit\`a Statale di Bergamo, 
24044--Dalmine (BG), Italy;
 
INFN--Sezione di Milano, Milan, Italy; \ and
 
DECOM, Faculty of Electrical Engineering (FEEC), State University at Campinas (UNICAMP), Campinas, Brazil.}

\vs{0.2 cm}

and
 
\vs{0.2 cm}

{\bf Giovanni SALESI}
 
{\em Facolt\`a di Ingegneria, Universit\`a Statale di Bergamo, 
24044--Dalmine (BG), Italy; \ and
 
INFN--Sezione di Milano, Milan, Italy.}}

\end{center}
 
\vs{1.0cm}

\centerline {\bf Abstract }

{\small {In previous papers we investigated the classical theory of Barut and Zanghi (BZ) for the electron spin [who interpreted the Zitterbewegung (zbw) motion as an internal motion along helical paths], and its ``quantum'' version, just by using the
language of Clifford algebras; and, in so doing, we ended with 
a new non-linear Dirac-like 
equation (NDE). \ We want to re-address here the whole subject,
and extend it, by translating it however
into the ordinary tensorial language, within the frame of a first 
quantization formalism. In particular, we re-derive here the
NDE for the electron field, and show it to be associated with a new
conserved probability current (which allows us to work out a ``quantum 
probabilistic" interpretation of the NDE).  \ Incidentally, the Dirac
equation results from the former by averaging over a zbw cycle.\hfill\break
\h Afterward, we derive an equation of motion for the 4-velocity field, which allows us to regard the electron as an extended-like object with a classically intelligible internal structure.\hfill\break
\h We carefully study the solutions of the NDE; with special attention to those implying (at the classical limit) light-like helical motions, since they
appear to be the most adequate solutions for the electron description from the kinematical and physical points of view, and do cope with the 
electromagnetic properties of the electron.\hfill\break
\h At last we introduce a natural generalization of our approach, for the case in which an external electromagnetic potential $A^\mu$ is present; \  it happens to be based on a new system of five first-order
differential field equations.}}

\
 
\

\

{\bf 1. -- Premise}\\

In previous work, we investigated the classical 
Barut--Zanghi (BZ) theory$^{[1]}$ for the
electron spin, which involves internal Zitterbewegung (zbw) motions along cylindrical helices, by using the powerful
language of Clifford algebras$^{[2]}$.  The  ``quantum" version of such approach lead us$^{[2]}$, in particular, to a new, non-linear, Dirac-like equation (NDE). 

\h In this work we implement a first quantization of the
BZ theory for the free electron, by adopting however the {\em ordinary} 
tensorial language. In so doing, we 
re-derive the NDE for the electron quantum field. Such equation involves {\em a 
new probability current} $J^\mu$, which is shown to be a conserved 
quantity, endowed (except in the ``trivial" case corresponding to the standard 
Dirac theory) with  a zbw motion having the typical frequency $\om=2m$, where $m$ is the electron mass.

\h Then, we derive a new equation of motion for the field velocity
$v^\mu$ quite different from that of Dirac's theory. 
Indeed, it allows us referring to realistic internal motions 
endowed with an intuitive kinematical meaning. The consequence is that
our electron results to be an extended-like particle, with a
classically intelligible internal structure. \ \ Afterwards, we write down the
general solution of the NDE, which appears to be a superposition of
plane waves with positive and negative frequencies; in the present
theory such superposition results to entail always a positive field energy, both for particles and for antiparticles.

\h We analyse the kinematical (zbw) structure of $v^\mu$
in the center-of-mass frame as a function of the initial spinor
$\psi(0)$. \ After having shown that, for extended-like particles like our electron, quantity $v^2 \equiv v_\mu v^\mu$ in general is {\em not} 
constant in time (quite differently from the scalar particle case, in which  $v^2=1$
is always constant), we look for the particular NDE solutions that on the contrary imply
a constant $v^2$; and we find this case to correspond to a
circular, uniform zbw motion with $\om=2 m$ and orbital radius $R=
|\vbf|/ 2m$ (where $|\vbf|$, and therefore $R$, depend on
the particular NDE solution considered). \ Even more, the
simple requirement of a uniform (classical) motion seems to play
the role of the ordinary quantization conditions for  the
$z$-component of the field spin $s$; \ namely,  $v^2=$ constant does imply
$s_z = \pm {1 \over 2}$, that is to say, the electron polarization. \  
We also examine, then, the oscillating linear motions associated with
unpolarized electrons.

\h Special attention is devoted to the light-like $(v^2=0)$ helical motions, either clockwise or anti-clockwise, for which the orbital radius results
to be equal to half a Compton wavelength. In fact, as already mentioned, such motions appear to
be the most adequate for the electron description, both kinematically
and physically, and correspond to the electromagnetic properties
of the electron, such as Coulomb field and intrinsic magnetic
moment.\\

{\bf 2. -- Introduction}\\
 
\h Attempts to put forth classical theories for the electron
are known since almost a century.$^{[3-6]}$  For instance, 
Schr\"{o}dinger's suggestion$^{[4]}$ that the electron spin was related to the
Zitterbewegung (internal) motion did originate a large amount of
subsequent work, including Pauli's. \ In ref.[7] one can find, for 
instance,  even the proposal
of models with clockwise and anti-clockwise ``inner motions"
as classical analogues of quantum relativistic spinning particles and
antiparticles, respectively.  \  Most of the cited works did not pay 
attention, however, to the
transition from the classical theory to its quantum version, i.e., to
finding out a wave equation whose classical limit be consistent with
the assumed model.
 
\h Among the approaches in which on the contrary a passage to the
quantum version was performed, let us quote the papers by Barut and
Pavsic.$^{[1,8]}$ \ Namely, those authors first considered a classical
lagrangian$^{[1]}$ $\Le$ describing (internal) helical motions; and second ---after having passed to the hamiltonian formalism---
they constructed a quantum version of their 
model and thus derived (in their opinion) the Dirac
equation. Such results are really quite interesting and stimulating. We
show below, however, that from the BZ
theory one does naturally derive a non-linear Dirac-like
equation,$^{[2]}$ rather than the Dirac (eigenvalue) equation itself, whose
solutions are only a subset of the former's.
 
\h Many further results have been met,$^{[2,9]}$, as we were saying,  
by using the powerful Clifford algebra language.$^{[10]}$ \ Due to their general 
interest and importance, we reformulate here  
the results appeared in refs.[2] by the {\em ordinary} tensorial language:
This will allow us to show more plainly and clearly, and more easily outline, their geometrical
and kinematical significance, as well as their field theoretical
implications. In so doing, also we shall further
develop such a theory; for instance: \ (i) by showing the
strict correlation existing between electron polarization and zbw
kinematics (in the case of both time-like and --with particular attention, as we know---
light-like internal helical motions);  and \ (ii) by forwarding a probabilistic
interpretation of the NDE current (after having shown it to be
always conserved).\\
 
{\bf 3. -- The classical Barut--Zanghi (BZ) theory}\\
 
\h The existence of an ``internal" motion (inside the electron) is denounced,
besides by the presence of spin, by the
remarkable fact that, according to the standard Dirac theory, the
electron four-impulse $p^\mu$ is {\em not} parallel to the four-velocity: \
$v^\mu \neq p^\mu/m$; \ moreover, while \ $[{p}^\mu,
{H}]=0$ \ so that $p^\mu$ is a conserved quantity, \ on the contrary
$v^\mu$ is {\em not} a constant of the motion: \ $[{v}^\mu, {H}]\neq
0$. \ Let us recall that indeed, if $\psi$ is a solution of Dirac
equation, only the ``quantum average" (or probability
current) \ $< v^\mu > \equiv \psi^{\dag} {v}^\mu \psi \equiv
\psi^{\dag} \ga^0 \ga^\mu \psi\equiv \ov{\psi} \ga^\mu \psi$ \ is constant in time
(and space), since it equals $p^\mu/m$. 

\h This suggests, incidentally, that  to
describe a spinning particle at least four independent canonically
conjugate variables are necessary; for example:
\[
x^\mu \, , \ p^\mu \, ; \ v^\mu \, , \ {{\scriptp}}^\mu \: ,
\]
or alternatively (as we are going to see):
\[
x^\mu \, , \ p^\mu \, ; \ z \, , \ \ov{z}
\]
where $z$ and $\ov{z}\equiv z^{\dag} \ga^0$ are ordinary 
${\rm {\C}}^4$--bispinors.
 
\h In the BZ theory,$^{[1]}$ the classical electron was actually
characterized, besides by the usual pair of conjugate variables
$(x^\mu, p^\mu)$, by a second pair of conjugate classical {\em spinorial}
variables $(z, \ov{z})$, representing internal degrees of freedom,
which are functions of the proper time $\tau$  of the electron global
center-of-mass (CM) system. The CM frame is the one in which at every instant of time it is
$\impulse = 0$ (but in which, for spinning particles, $\vbf \equiv 
\dot{\xbf}$ is {\em not} zero, in general!).  
 \ Barut and Zanghi, then,
introduced a classical lagrangian that in the free case ($A_\mu = 0$) 
writes $[c=1]$
\bb
\Le=\frac{1}{2} i \la (\dot{\ov{z}}z - {\ov{z}}\dot{z}) +
p_\mu(\dot{x}^\mu - \ov{z}\ga^\mu z)
\ee
where $\la$ has the dimension of an action. 
[The extension of this lagrangian to the case with
external fields, $A^\mu \neq 0$, has been treated elsewhere].
One of the consequent (four) motion equations yields directly the particle four-velocity:
$$
\dot{x}^\mu \equiv v^\mu = \ov{z} \ga^\mu z \; .
\eqno {\rm(1')} $$
 
We are not writing down explicitly the spinorial indices of $\ov{z}$ and $z$.

\h Let us explicitly notice that, in the classical theories for 
elementary spinning particles, it is convenient$^{[11]}$ to split the
motion variables as follows
\bb
x^\mu \equiv \xi^\mu + X^\mu \; ; \ \  \ v^\mu = w^\mu + V^\mu \ ,
\ee
where $\xi^\mu$ and $w^\mu\equiv \dot{\xi}^\mu$ describe the 
{\em translational, external} or {\em drift} motion, i.e. the motion of the  
CM, whilst $X^\mu$ and $V^\mu \equiv \dot{X}^\mu$ describe the {\em internal}
or {\em spin} motion.

\h From eq.(1) one can see$^{[1]}$ that also
\bb
H \equiv p_\mu v^\mu = p_\mu \ov{z} \ga^\mu z
\ee
is a constant of the motion (and precisely it is the CMF energy); being $H$, 
as one may easily check, the BZ
hamiltonian in the CMF, we can suitably set $H=m$, quantity 
$m$ being the particle rest-mass. In this way, incidentally,
we obtain, even for spinning particles, the ordinary 
relativistic constraint (usually met for scalar particles):
$$
p_\mu  v^\mu= m \ .
\eqno {\rm(3')} $$
 
\h The general solution of the equations of
motion corresponding to lagrangian (1), with $\la = -1$ \ [and $\hbar = 1$], is:
$$
z(\tau)=[\cos (m\tau)- i \frac{p_\mu \ga^\mu}{m}\sin (m\tau)]z(0) \  ,
\eqno {\rm(4 a)} $$
$$
\ov{z}(\tau)=\ov{z}(0)[\cos (m\tau)+ i \frac{p_\mu \ga^\mu}{m}
\sin (m\tau)] \  ,
\eqno {\rm(4 b)} $$
 
with $p^\mu=$ constant; \ $p^2=m^2$; \ and finally:   
$$
\dot{x}^\mu\equiv v^\mu=\frac{p^\mu}{m}+[\dot{x}^\mu(0)-\frac{p^\mu}{m}]
\cos(2m\tau)+\frac{\ddot{x}^\mu}{2m}(0)\sin (2m \tau) \ .
\eqno {\rm(4 c)} $$
 
This general solution exhibits the classical analogue of the phenomenon known 
as Zitterbewegung: in fact, the velocity $v^\mu$ contains the
(expected) term $p^\mu/m$ plus a term describing an oscillating
motion with the characteristic zbw frequency $\om=2m$. \ The velocity
of the CM will is given of course by $w^\mu=p^\mu/m$. 

\h Notice that,
instead of adopting the variables $z$ and $\ov{z}$, we can work in terms of 
the spin variables, i.e., in terms of the set of dynamical variables
\[
x^\mu \, , \ p^\mu \, ; \ v^\mu \, , S^{\mu \nu}
\]
where
$$
S^{\mu \nu} \equiv {i \over 4} \, \ov{z} [\ga^\mu, \ga^\nu] z \; ;
\eqno {\rm(5a)} $$
 
then, we would get the following motion equations:
$$
\dot{p}^\mu=0 \ ; \ \ v^\mu=\dot{x}^\mu \ ; \ \ \dot{v}^\mu=4 S^{\mu
\nu}p_{\nu} \ ; \ \ \dot{S}^{\mu \nu}= v^\nu p^\mu - v^\mu p^\nu \ .
\eqno {\rm(5b)}  $$
 
\h By varying the action corresponding to $\Le$, one finds as
generator of space-time rotations the conserved quantity \ $J^{\mu
\nu}=L^{\mu\nu}+ S^{\mu\nu}$, \ where \ $L^{\mu\nu} \equiv x^\mu p^\nu - 
x^\nu p^\mu$ \ is the orbital
angular momentum tensor, and $S^{\mu\nu}$ is just the particle spin tensor: 
 \ so that \ 
${\dot{J}}^{\mu\nu}=0$ \ implies \ $\dot{L}^{\mu\nu}=-\dot{S}^{\mu\nu}$. 

\h Let us explicitly observe that the general solution (4c)
represents a helical motion in the ordinary 3-space: a result that has been met
also by means of other,$^{\#1}$ alternative approaches.$^{[12,13]}$\\

\footnotetext{$^{\#1}$ Alternative approaches to the kinematical description
of the electron spin have been proposed, e.g., by Pavsic and Barut in 
refs.[12,13]. \ In connection with Pavsic's approach,$^{[12]}$  we
would like here to mention that the classical angular momentum was defined 
therein as \ $\sbf \equiv 2 \beta \vbf \wedge \abf / {\sqrt{1 - {\vbf}^2}}$, \
whilst in the BZ theory it is \ $\sbf \equiv \rbf \wedge m \wbf$, \ where 
$\abf \equiv \dot{\vbf}$. Both quantities $\sbf$ result to be parallel 
to~$\impulse$.}

{\bf 4. --  Field theory of the extended--like electron}\\
 
\h The natural way of ``quantizing" lagrangian (1) is that of
reinterpreting the classical spinors $z$ and $\ov{z}$ as Dirac field spinors, 
say $\psi$
and $\ov{\psi}$ [in the following the Dirac spinors will be merely called
``spinors", instead of bi-spinors, for simplicity's sake]:
\[
z \rig \psi \ \ ; \ \ \ov{z} \rig \ov{\psi} \ \ ;
\]
which will lead us below to the conserved probability current \ 
$J^\mu= m \: \ov{\psi}\ga^\mu \psi / {p^0}$. \ Recall that here the 
operators $(x^\mu, p^\mu; \psi, \ov{\psi})$ are field variables;
for instance,$^{[2]}$ \ $\psi = \psi(x^\mu); \;\; \ov{\psi} = 
\ov{\psi}(x^\mu)$. \  
Thus, the quantum version of eq.(1) is the field lagrangian
\setcounter{equation}{5}
\bb
\Le = \frac{i}{2} \la  (\dot{\ov{\psi}}\psi - \ov{\psi}\dot{\psi})
+ p_\mu (\dot{x}^\mu - \ov{\psi}\ga^\mu \psi)
\ee
that refers (so as in the classical case) to free electrons with {\em fixed}
impulse $p^\mu$. \ The four Euler--Lagrange equations, with $-\la=\hbar
=1$, yield the following motion equations: 
 
\
 
$\hfill{\dis\left\{\begin{array}{l}
\dot{\psi}+ i p_\mu \ga^\mu \psi=0\\

\dot{x}^\mu= \ov{\psi}\ga^\mu \psi\\

\dot{p}=0 \; ,
\end{array}\right.}
\hfill{\dis\begin{array}{r}
(7{\rm a}) \\ (7{\rm b}) \\ (7{\rm c}) \end{array}}$
 
\
 
besides the hermitian adjoint of eq.(7a), holding for $\ov{\psi}= \psi^ + 
\ga^0$. \ In eqs.(7),  the invariant time $\tau$ is still the CMF proper time,
and it is often put \ $p_\mu \ga^\mu \equiv \pp$.
 
\h We can pass to the hamiltonian formalism by introducing the field 
hamiltonian        
corresponding to the energy in the CMF:
\setcounter{equation}{7}
\bb
H=p_\mu \ov{\psi} \ga^\mu \psi \; ,
\ee
which easily yields {\em the noticeable equation\/}$^{\#2}$
\footnotetext{$^{\#2}$ In refs.[12] it was moreover assumed that $p_\mu
\dot{x}^\mu=m$, which actually does imply the very general relation
$p_\mu \ov{\psi}\ga^\mu \psi=m$, but {\em not} the Dirac equation 
$p_\mu \ga^\mu \psi = m \psi$, as claimed therein; these two equations 
in general are not equivalent.}   
\bb
p_\mu \ov{\psi} \ga^\mu \psi=m
\ee
 
\h This {\em non-linear} equation, satisfied by all the solutions of eqs.(7),
is very probably the simplest non-linear {\em Dirac--like} equation.$^{[14]}$ 
{\em Another} non-linear Dirac--like equation,
quite equivalent to eq.(7a) but employing the generic coordinates
$x^\mu$ and no longer the CMF proper time, was anticipated in the first one of
refs.[2]; \ it is obtained by inserting the identity
\bb
\frac{\drm}{\drm \tau} \equiv \frac{\drm x^\mu}{\drm \tau} \frac{\pa}{\pa
x^\mu}\equiv {\dot{x}}^\mu \pa_\mu  
\ee
into eq.(7a). In fact, one gets \ \  
$i \dot{x}^\mu \pa_\mu \psi = p_\mu \ga^\mu \psi$, \ \ 
and, since $\dot{x}^\mu = \ov{\psi} \ga^\mu \psi$ because of eq.(7b), one
arrives at the important equation:  
\bb
i \ov{\psi} \ga^\mu \psi \pa_\mu \psi= p_\mu \ga^\mu \psi \ .
\ee
 
A more general equation, in which $p_\mu \ga^\mu$ is replaced by $m$,
\[
i \ov{\psi} \ga^\mu \psi \pa_\mu \psi=  m \psi \ ,
\]
can be easily obtained (cf. the last one of refs.[2]) by releasing the  
fixed--$p^\mu$ condition.

\h Let us notice that, differently from eqs.(6)--(7), equation (11) can
be valid a priori even for massless spin $1 \over 2$ particles, since
the CMF proper time does not enter it any longer.
 
\h The remarkable non-linear equation (11) corresponds to the whole
system of eqs.(7): quantizing the BZ theory, therefore, does not
lead to the Dirac equation, but rather to the non-linear, Dirac-like
equation (11), that we call NDE. 

\h The eq.(11) might be even adopted in substitution
for the free Dirac (eigenvalue) equation, since it apparently admits a 
sensible classical limit [which describes an internal periodic motion with
frequency $2 m$, implying the existence of an intrinsic angular momentum
tensor: the ``spin tensor" $S^{\mu \nu}$ of eq.(5a)]. Moreover, eq.(11)
admit assigning a natural, simple physical meaning to the negative
frequency waves.$^{[15]}$ \  In other words, eqs.(6)--(11) seem to allow us overcoming the wellknown problems related with physical meaning and time
evolution of the position operator $x^\mu$ and of the velocity
operator $\dot{x}^\mu$: we shall come back to this point. [We always 
indicate a variable and the corresponding operator by the same simbol]. \ 
In terms of field quantities (and no longer of operators), eq.(11) 
corresponds$^{[2]}$ to the four motion equations
$$
\dot{p}^\mu=0; \ \ v^\mu=\dot{x}^\mu; \ \ \dot{v}^\mu=4 S^{\mu\nu}p_\nu;
 \ \ \dot{S}^{\mu \nu} = v^\nu p^\mu -v^\mu p^\nu \; ,
\eqno {\rm(12a)} $$
in which now $v^\mu$ and the spin tensor $S^{\mu\nu}$ are the field 
quantities \ $v^\mu\equiv \ov{\psi} (x) \ga^\mu \psi (x)$ \ and 
$$
S^{\mu\nu}=\frac{i}{4}\ov{\psi}(x) [\ga^\mu, \ga^\nu]\psi (x) \; .
$$  
By deriving the third one of eqs.(12a), and using the first one of
them, we obtain
$$
\ddot{v}^\mu = 4\dot{S}^{\mu\nu} p_\nu \ ;
\eqno {\rm (12b)} $$ 
by substituting now the fourth one of eqs.(12a) into eq.(12b), and imposing
the previous constraints $p_\mu p^\mu=m^2, \ \ p_\mu v^\mu=m$, we get the 
time evolution of the field four-velocity:
\setcounter{equation}{12}
\bb
v^\mu= \frac{p^\mu}{m}-\frac{\ddot{v}^\mu}{4 m^2} \ .
\ee
Let us recall, for comparison, that the corresponding
equation {\em for the standard Dirac case}$^{[1-7]}$ was devoid of a
classical, realistic
meaning because of the known appearance of an imaginary unit $i$ in front of 
the acceleration:$^{[16]}$
$$
v^\mu = {\dis\frac{p^\mu}{m}- \frac{i}{2m}} \dot{v}^\mu \ .
\eqno {\rm (13')} $$
 
\h One can observe, incidentally, that by differentiating the relation
$p_\mu v^\mu = m = {\rm constant}$, one immediately gets that the (internal)
acceleration $\dot{v}^\mu\equiv \ddot{x}^\mu$ is orthogonal to the electron
impulse $p^\mu$ since $p_\mu \dot{v}^\mu=0$ at any instant. To conclude, let 
us recall that the Dirac electron has
no classically meaningful internal structure; on the contrary, our
electron, an {\em extended--like} particle, does possess an internal
structure, and internal motions which 
are all kinematically, geometrically acceptable and meaningful: As we are going to see.\\
 
{\bf 5. -- General solution of the new non-linear, Dirac--like equation (NDE),
and conservation of the probability current}\\
 
\h In a generic frame, the general solution of eq.(11) can be easily shown
to be the following [$\pp \equiv p_\mu \ga^\mu$]:
$$
\psi (x)=[{\frac{m-\pp}{2 m} \; e^{i p_\mu x^\mu} + 
\frac{m+\pp}{2 m} \;  e^{-i p_\mu x^\mu}}] \ \psi(0) \ ;
\eqno {\rm(14a)} $$
 
which, in the CMF, reduces to
$$
\psi (\tau)=[{\frac{1-\ga^0}{2} \; e^{i m \tau} + 
\frac{1+\ga^0}{2} \; e^{-i m\tau}}] \ \psi(0) \ ,
\eqno {\rm (14b)} $$
 
or, in a simpler form, to:
$$
\psi (\tau)=[\cos(m \tau)- i \ga^0 \sin (m\tau)] \ \psi (0) \ ,
\eqno {\rm (14c)} $$
 
quantity $\tau$ being the particle proper time, as above. \ Let us explicitly 
observe that, by introducing eq.(14a), or eq.(14b), into eq.(9),  one obtains 
that every solution of eq.(11) does correspond to the CMF field hamiltonian 
$H=m > 0$,  even if it appears (as expected in any theories with zbw) to be a 
suitable superposition of plane waves endowed with positive and negative
frequencies.$^{[15]}$ \ Notice that superposition (14a) is a
solution of eq.(11), due to the non-linearity of such an equation, {\em only} 
for suitable pairs of plane waves, with weights
\[
\frac{m \pm \pp}{2 m} = {\La}_\pm \; ,
\]
respectively, which are nothing but the usual projectors 
${\La}_+ \;\; ({\La}_-)$ over
the positive (negative) energy states of the standard Dirac equation.
In other words, the plane wave solution (for a fixed value of $p$) of the Dirac
eigenvalue equation $\pp \psi = m \psi$ is a particular case of the
general solution of eq.(11): namely, for either
\setcounter{equation}{14}
\bb
{\La}_+ \psi (0) = 0  \;\;\;\;\;\;\; {\rm or} \;\;\;\;\;\;\; {\La}_- \psi (0) 
= 0 \; .
\ee
 
Therefore, the solutions of the Dirac eigenvalue equation
are a {\em subset} of the set of solutions of our NDE. \  It is
worthwhile to repeat that, for each fixed $p$, the wave function
$\psi(x)$ describes both particles and antiparticles: all
corresponding however to positive {\em energies}, in agreement with the
reinterpretation forwarded in refs.[15] (as well as with the already 
mentioned fact that we can always choose $H = m > 0$).\\ 

\h We want now to study, eventually, the probability current
$J^\mu$ corresponding to the wave functions (14a,b,c). Let us define it as
follows:
\bb
J^\mu \equiv \frac{m}{{p^0}} \; \ov{\psi} \ga^\mu \psi
\ee
where the normalization factor \ $m/ {p^0}$ \ (the 3-volume $V$ being
assumed to be equal to 1, as usual; so that ${p^0} V \equiv
{p^0}$) is required to be such that the classical limit of $J^\mu$, that is
 \ $(m/{{p^0}}) \, v^\mu$, \ equals \ $(1; \vbf)$, \ like for the ordinary
probability currents. \ Notice also that sometimes in the literature $p^0$ 
is replaced by $\scrE$; and that $J^0\equiv 1$, which means that
we have one particle inside the unitary 3-volume $V=1$. \ This normalization
allows us to recover, in particular,  the Dirac current
$J^\mu_{\rm D}=p^{\mu}/ {p^0}$ when considering the (trivial) solutions,
without zbw, corresponding to relations (15).  
\h Actually, if we substitute 
quantity $\psi(x)$ given by eq.(14a) into eq.(16), we get
$$
J^\mu={\dis\frac{p^\mu}{{p^0}}}+E^\mu \cos(2 p_\mu x^\mu)+ H^\mu
\sin (2 p_\mu x^\mu) \; ,
\eqno {\rm (16')} $$
 
where
$$
E^\mu \equiv J^\mu(0)- p^\mu/ {p^0} \; ; \ \ \ H^\mu\equiv \dot{J}(0) 
/ 2 m \; .
\eqno {\rm (16'')} $$
 
If we now impose conditions (15), we have $E^\mu=H^\mu=0$ and get therefore
the Dirac current $J^\mu= J_{\rm D}^\mu= \mbox{constant}=p^\mu/
{p^0}$.
Let us notice too that the normalization factor $\sqrt{m/
{p^0}}$ cannot be inserted into $\psi$ and $\ov{\psi}$, as it
would seem convenient, because of the non-linearity of
eq.(11) and/or of constraint (9).
 
\h Since $p_\mu E^\mu \equiv p_\mu J^\mu (0)-p_\mu p^\mu/ {p^0} = m^2/
{p^0} - m^2/ {p^0} = 0$ (where we used eq.(9) for $x=0$) and
since $p_\mu H^\mu \equiv p_\mu \dot{J}^\mu (0)/ 2m=0$ obtained
deriving both members of eq.(9) ---note incidentally that both
$E^\mu$ and $H^\mu$ are orthogonal to $p^\mu$--- it follows that
$$
\pa_\mu J^\mu=2 p_\mu H^\mu \cos (2 p x)- 2 p_\mu E^\mu \sin (2 p
x)=0 \ .
\eqno {\rm (16''')} $$
 
\h We may conclude, with reference to equation (11), that our
current $J^\mu$ is {\em conserved}: We are therefore allowed to adopt the
usual probabilistic interpretation of fields $\psi, \ov{\psi}$.
Equation (16') does clearly show that the conserved current $J^\mu$, as
well as its classical limit $(m / {p^0}) v^\mu$ \ [see eq.(4c)],
are endowed with a Zitterbewegung--type motion:  precisely,
with an oscillating motion having the CMF frequency  $\Om= 2
m \simeq 10^{21} {\rm s}^{-1}$ and period $T = \pi / m \simeq
10^{-20}$s (we may call $\Om$ and $T$ the zbw frequency and period,
respectively).
 
\h From eq.(16') one can immediately verify that in general 
\[ 
J^\mu \neq p^\mu/ {p^0} \; , \ \ \ \ J^\mu\equiv J^\mu(x) \; ; 
\]
whilst the Dirac current $J_{\rm D}^\mu$ for the free
electron with fixed $p$, as already mentioned, is constant:
\[
J^\mu_{\rm D}= p^\mu / {p^0} = \ {\rm constant} \; ,
\]
corresponding to {\em no} zbw. \ 
In other words, our current behaves differently from Dirac's, 
even if both of them obey$^{\#3}$ the constraint [cf. eq.(9)]
\footnotetext{$^{\#3}$ In the Dirac case, this is obtained by getting 
from the ordinary Dirac equation, \ 
$p_\mu \ga^\mu \psi_{\rm D}=m \psi_{\rm D}$, \ the non-linear constraint $p_\mu
\ov{\psi}_{\rm D} \ga^\mu \psi_{\rm D}=m\ov{\psi}_{\rm D} \psi_{\rm D}$, and
therefore by
replacing $\ov{\psi}_{\rm D} \psi_{\rm D}$ by $m/{p^0}$, consistently with the
ordinary normalization $\psi_{\rm D} = e^{-ipx}u_p / \sqrt{2 {p^0}}$, with 
${\ov{u}}_p u_p = 2m$.}
\[
p_\mu J^\mu= p_\mu J^\mu_{\rm D} = m^2/ {p^0} \ .
\] 
\h It's noticeable, moreover, that our current $J^\mu$ goes into the
Dirac one, not only in the no-zbw case of eq.(15), but also when
averaging it (in time) over a zbw period:
\bb
< J^\mu >_{\rm zbw} = \frac{p^\mu}{{p^0}} \equiv J^\mu_{\rm D} \ .
\ee
 
\h In the next section we study the kinematical zbw structure
of $J^\mu$, relative to some given solutions of the NDE: Let us stress
that this structure is the same for both  $J^\mu$ and  its
classical limit ${m v^\mu} / {p^0}$ (and that in the CMF
it is  ${m v^\mu} / {p^0}  \equiv v^\mu$). That
is to say, the probability current stream-lines correspond just to the
classical world-lines of a pointlike particle, in agreement with the
Correspondence principle. \              
Consequently, we can 
study the kinematical features of our $J^\mu$ by means of
the analysis of the time evolution of the four-velocity $v^\mu$.\\
 
{\bf 6. -- Uniform motion solutions of the NDE}\\
 
\h Before examining the solutions, let us stress that we
do use the term  {\em electron\/}$^{\#4}$
\footnotetext{$^{\#4}$ Usually we speak about electrons, but this theory 
could be applied to all leptons.} 
to indicate the
whole spinning system associated with the geometrical
center of  the helical trajectories (which corresponds to the center
of the electron Coulomb field$^{[17]}$). let us repeat that this
geometrical center is always at rest in the CMF, i.e., in the
frame where $\impulse$ and $\wbf$ (but {\em not}
$\vbf \equiv \Vbf$) vanish [cf. eqs.(2)]. On the contrary, we shall
call $\cal Q$ the pointlike object moving along the helix; we shall refer
to it as to the electron ``constituent",  and to its (internal) movement as to
a ``sub-microscopic" motion.
 
\h We need first of all to make explicit the kinematical definition of
$v^\mu$, rather different from the common (scalar particle)
one.  In fact, from
the very definition of $v^\mu$, one obtains
\[
v^\mu\equiv \drm x^\mu/ \drm \tau \equiv (\drm t/ \drm \tau; 
\drm \xbf / \drm \tau)
\equiv ({\frac{\drm t}{\drm \tau}}; \frac{\drm \xbf}{\drm t} \; 
\frac{\drm t}{\drm \tau})
\]
\bb
=(1/ \sqrt{1- \wbf^2}; \;\; \ubf / \sqrt{1- \wbf^2}) \; , \;\;\;\;\;\; \quad
[\ubf \equiv \drm \xbf / \drm t]
\ee
[where, let us recall,  $\wbf = \impulse/ m$ is the velocity of the CM in the chosen
reference frame (that is, in the frame in which the quantities $x^\mu$ are 
measured)]. \ Below, it will be convenient to choose as reference frame the
CMF  (even if quantities as $v^2 \equiv v_\mu v^\mu$
are frame invariant);  so that [cf. definition (2)]:
\bb
v^\mu_{\rm CM}= V^\mu \equiv (1; \Vbf) \ ,
\ee
wherefrom one deduces for the speed $|\Vbf|$
of the internal motion (i.e., for the zbw speed) the new conditions:
\[
0 < V^2 (\tau) < 1 \;\;\; \Lef \;\;\; 0 < \Vbf^2 (\tau) < 1 \;\;\;\;\;\;
\quad\mbox{(time-like)}
\]
$$
V^2(\tau)=0 \quad \Lef \quad \Vbf^2 (\tau) = 1 \;\;\;\;\; \ \ \ \ \ \ \ \ 
 \quad\mbox{(light-like)}
\eqno {\rm (19')} $$ 
\[
V^2 (\tau) < 0 \quad \Lef \quad \Vbf^2 (\tau) > 1 \;\;\;\;\; \ \ \ \ \ \ \ \ 
\quad\mbox{(space-like)} \ ,
\]
where $V^2 = v^2$. \  Notice that, in general, the value of $\Vbf^2$ does vary
with $\tau$; except in special cases (e.g., the case of polarized particles: 
as we shall see).
 
\h Our NDE in the CMF (where, let us
remember, $J^\mu \equiv v^\mu$) can be written as
\bb
(\ov{\psi} \ga^\mu \psi) i \pa_\mu \psi= m \ga^0 \psi
\ee
whose general solution is eq.(14b) or, equivalently, eq.(14c). \
In correspondence to it, we also have [due to eq.(16')] that
$$
J^\mu= p^\mu /m + E^\mu \cos (2m \tau)+ H^\mu\sin (2m \tau)\equiv V^\mu
\eqno {\rm (21a)} $$
$$
V^2 \equiv J^2= 1+ E^2 \cos^2 (2m \tau)+ H^2 \sin^2(2m \tau)+
2 E_\mu H^\mu \sin (2m \tau) \cos (2m \tau) \ .
\eqno {\rm(21b)} $$
 
\h Let us select the solutions $\psi $ of eq.(20) corresponding to
constant $V^2$ and $A^2$, where $A^{\mu} \equiv {\drm V^\mu} / {\drm \tau} 
\equiv (0; \Abf)$, quantity $V^\mu \equiv (1; \Vbf)$ being the zbw velocity. 
Therefore, we shall suppose in the present frame that quantities
\[
V^2= 1- \Vbf^2 \ \ ; \ \ A^2= -\Abf^2
\]
are constant in time:
\setcounter{equation}{21}
\bb
V^2 \ = \ {\rm constant} \; ; \ \ \ \  A^2 \ = \ {\rm constant} \; ,
\ee
so that $\Vbf^2$ and $\Abf^2$ are constant in time too. 
(Notice, incidentally, that we are dealing
exclusively with the internal motion, in the CMF; thus, our results are quite
independent of the global 3-impulse $\impulse$ and hold both in the
relativistic and in the non- relativistic case). \ The requirements (22),
inserted into eq.(21b), yield the following constraint

$\hfill{\dis\left\{\begin{array}{l}
E^2= H^2\\
E_\mu H^\mu=0 \; .\\
\end{array}\right.}
\hfill{\dis\begin{array}{r}
(23{\rm a}) \\ (23{\rm b}) \end{array}}$
 
\
 
\h Constraints (23) are necessary and sufficient (initial) conditions
to get a circular uniform motion (the only uniform
and finite motion conceivable in the CMF). \ Since
both $E$ and $H$ do not depend on the time $\tau$, also eqs.(23) hold
for any time instant. \ In the euclidean 3-dimensional space, 
and at any instant of time, constraints (23) read:
 
\

$\hfill{\dis\left\{\begin{array}{l}
\Abf^2= 4 m^2 \Vbf^2\\
\Vbf \cdot \Abf = 0\\
\end{array}\right.}
\hfill{\dis\begin{array}{r}
(24{\rm a}) \\ (24{\rm b}) \end{array}}$
 
\
 
which correspond to a uniform circular motion with radius
$$
R= |\Vbf|/ 2 m \ .
\eqno {\rm (24c)} $$
 
\h Quantity $R$ is the ``Zitterbewegung radius"; the zbw frequency
was already found to be always $\Om= 2m$. \ Quantities $E, H$ 
(with ${p^0} = m$) given in
eqs.($16''$) are functions of the initial spinors $\psi (0),
\ov{\psi} (0)$. Bearing in mind that in the CMF $v^0=1$ [cf.
eq.(19)] and therefore $\ov{\psi} \ga^0 \psi=1$ (which, incidentally,
implies the CMF normalization $\psi^{\dag} \psi=1$), one gets
$$
E^\mu \equiv {\dis-\frac{p^\mu}{m}}+\ov{\psi} (0) \ga^\mu \psi(0)=
(0; \; \ov{\psi} (0) \vec{\ga} \psi (0)) \ .
\eqno {\rm (24'a)} $$
 
By substituting into the second one of eqs.($16''$) the expression (14c) for the
general solution of the NDE,  we finally have
$$
H^\mu = i \ov{\psi}(0) [\ga^0 \ga^\mu- g^{0\mu}] \psi (0)=(0; \;
i \ov{\psi}(0) \vec{\al} \psi(0)) \; ,
\eqno {\rm (24'b)} $$
 
where $\vec{\ga} \equiv (\ga^1, \ga^2, \ga^3)$, while $\vec{\al}
\equiv \ga^0 \vec{\ga}$ and $g^{\mu \nu}$ is the metric. Therefore,
conditions (23) or (24) can be written  in spinorial form, for any
time $\tau$, as follows
 
\

$\hfill{\dis\left\{\begin{array}{l}
(\ov{\psi} \vec{\ga} \psi)^2= -(\ov{\psi} \vec{\al} \psi)^2\\
(\ov{\psi}\vec{\ga}\psi) \cdot (\ov{\psi} \vec{\al}\psi)=0 \; .
\end{array}\right.}
\hfill{\dis\begin{array}{r}
(25{\rm a}) \\ (25{\rm b}) \end{array}}$

\
 
\h At this point, let us show that this classical
uniform circular motion, occurring around the  $z$-axis (which in the
CMF can be chosen arbitrarily, while in a generic
frame is parallel to the global three-impulse $\impulse$, as we shall see), 
does just correspond to the case of {\em polarized} particles
with $s_z = \pm {1 \over 2}$. \ It may be interesting to notice, once more,
that in this case the classical requirements (23) or (24) ---namely,
the uniform motion conditions--- play the role of the ordinary
quantization conditions \ $s_z = \pm {1 \over 2}$. \ Now, let us first 
observe [cf. eq.(5a)] that in the CMF the $z$-component of the spin vector
\[
s_z= S^{12} \equiv \frac{i}{4} \ov{\psi} (\tau)[\ga^1, \ga^2] \psi
(\tau) 
\]
can actually be shown to be 
a constant of the motion. \ 
In fact, by easy calculations on eq.(14c), one finds $s_z$
to be independent of $\tau$:
\setcounter{equation}{25}
\bb
s_z (\tau)= \frac{i}{2} \ov{\psi}(0) \ga^1 \ga^2 \psi (0)= \frac{1}{2}
\ov{\psi}(0) \Sig_z \psi (0) \; ;
\ee
where $\vec{\Sig}$ is the spin operator, that in the standard
representation reads
\[
\vec{\Sig} \equiv
\left(\begin{array}{ll}
\vec{\sig} & 0\\
\\
0 & \vec{\sig} 
\end{array}\right)
\]
 
quantities $\sig_x, \sig_y, \sig_z$ being the well-known  
Pauli $2 \times 2$ matrices.
 
\h Then, it is straightforward to realize that the most general
spinors $\psi(0)$ satisfying the conditions
$$
s_x = s_y=0
\eqno {\rm(27a)} $$
$$
s_z = {\frac{1}{2}\ov{\psi}(0)\Sig_z \psi(0)} = \pm {1 \over 2}
\eqno {\rm(27b)} $$
 
are (in the standard representation) of the form
$$
\psi_{(+)}^{\rm T} (0)= ( \ a \ 0 \ | \ 0 \ d \ )
\eqno {\rm(28a)} $$
$$
\psi_{(-)}^{\rm T} (0)= ( \ 0 \ b \ | \ c \ 0 \ ) \; , 
\eqno {\rm(28b)} $$
 
respectively; and obey in the CMF the normalization constraint $\psi^{\dag}
\psi=1$. \ [It can be easily shown that, for generic initial conditions, 
it is \ $-{1 \over 2} \leq s_z \leq {1 \over 2}$]. \ In
eqs.(28) we separated the first two from the second two
components, bearing in mind that in the standard Dirac theory (and in the CMF)
they correspond to the positive and negative
frequencies, respectively. \ With regard to this, let us observe that the
``negative-frequency"
components $c$ and $d$ do {\em not} vanish at the non-relativistic limit [since,
let us repeat, it is $\impulse = 0$ in the CMF]; but the field
hamiltonian $H$ is nevertheless positive and
equal to $m$, as already stressed.
 
\h Now, from relation (28a) we are able to deduce that (with $*
\equiv$ complex conjugation)
\begae
< \vec{\ga} > &\equiv& \ov{\psi} \vec{\ga} \psi= 2( \Real [a^* d], + \Imag
[a^* d], 0)\\
< \vec{\al} > &\equiv& \ov{\psi} \vec{\al} \psi= 2i( \Imag [a^* d], - \Real
[a^* d], 0)
\egae
and analogously, from eq.(28b), that
\begae
< \vec{\ga} > &\equiv& \ov{\psi} \vec{\ga} \psi= 2( \Real [b^* c], - \Imag
[b^* c], 0)\\
< \vec{\al} > &\equiv& \ov{\psi} \vec{\al} \psi= 2i( \Imag [b^* c], + \Real
[b^* c], 0) \; ,
\egae
which just imply relations (25):
 
\[
\left\{\begin{array}{l}
< \vec{\ga} >^2 = - < \vec{\al} >^2\\
\\
< \vec{\ga} > \cdot < \vec{\al} > =0 \ .
\end{array}\right.
\]
 
\h In conclusion, the (circular) polarization conditions eqs.(27)
imply the internal zbw motion to be uniform and circular ($V^2=$
constant; $A^2=$ constant); eqs.(27), in other words, do imply
that $s_z$ is conserved and quantized, at the same time.
 
\h Notice that, when passing from the CMF to a generic frame,
eq.(27) transforms into
\setcounter{equation}{28}
\bb
\la \equiv {1 \over 2} \ov{\psi} \frac{\vec{\Sig} \cdot
\impulse}{|\impulse|} \psi \; = \; \pm {1 \over 2} \; = \; {\rm constant} \ .
\ee
 
Therefore, to get a uniform motion around the
$\impulse$-direction [cf. equations (4c) or (16')], we have to request
that the helicity $\la$ be constant (over space and time), and
quantized in the ordinary way, i.e., $\la = {1 \over 2}$. \ We shall 
come back to the
question of the double sign $\pm {1 \over 2}$ in the case of the light-like 
helical
trajectories; here, for simplicity, let us confine to the $+$ sign.
 
\h It may be interesting, now, to calculate $|\Vbf|$ as a function
of the spinor components $a$ and $d$. \ With reference to eq.(28a), since
$\psi^{\dag} \psi \equiv |a|^2 + |d|^2=1$, we obtain (for the $s_z= + 
{1 \over 2}$
case):
$$
\Vbf^2 \equiv < \vec{\ga} >^2= 4| a^* d|^2=4 |a|^2 \; (1-|a|^2)
\eqno {\rm(30a)} $$
$$
\Abf^2 \equiv (2 i m < \vec{\al} >)^2= 4m^2 \Vbf^2=16 m^2 |a|^2
 \; (1-|a|^2)\ ,
\eqno {\rm(30b)} $$
and therefore the normalization (valid now in any frame, at any time) 
$$
\ov{\psi} \psi = \sqrt{1 - \Vbf^2} \ ,
\eqno {\rm(30c)} $$
which show that to the same speed and acceleration there correspond two spinors
$\psi(0)$, related by an exchange of $a$ with $d$. \ From eq.(30a)
we derive also that, as $0 \leq |a| \leq 1$, it is:
$$
0 \leq \Vbf^2 \leq 1 \; ; \ \ \ 0 \leq \ov{\psi} \psi \leq 1 \; .
\eqno {\rm(30d)} $$
 
Correspondingly, from eq.(24c) we obtain for the zbw radius \ 
$0 \leq R \leq {1 \over 2} m$. \\  
\h The second one of eqs.(30d) is a {\em new}, rather interesting 
(normalization)
boundary condition. \ From eq.(30c) one can easily see that: \ (i) for
$\Vbf^{2} = 0$ (no zbw) we have $\ov{\psi}\psi = 1 $ and $\psi$ is a
``Dirac spinor''; \ (ii) for $\Vbf^{2} = 1$ (light-like zbw) we have
$\ov{\psi}\psi = 0$ and $\psi$ is a ``Majorana spinor''; \ 
(iii) for $0 < \Vbf^{2} < 1$ 
we meet, instead, spinors with properties ``intermediate'' between
the Dirac and the Majorana ones.\\
\h As an example, let us write down the ``Caldirola$^{[18]}$ solution"
$\psi(0)$, corresponding to the zbw speed $\sqrt{3}/2$ and to the zbw
radius $\sqrt{3}/4 m$,  and yielding correct values for the zero
and first order contributions to the electron magnetic moment (for
simplicity we chose $a$ and $d$ real):
\[
\psi^{\rm T} (0)= (\frac{1}{2} \ \ 0 \ | \ 0 \ \frac{\sqrt{3}}{2}) \ ;
\]
as well as that got by interchanging $1 / 2$ and $\sqrt{3}/2$.
 
\h The ``Dirac" case,$^{[2,9]}$ corresponding to $\Vbf^2=
\Abf^2 = 0$, that is to say, corresponding  to no zbw internal motion, is merely represented (apart from phase factors) by the spinors
\setcounter{equation}{30}
\bb
\psi^{\rm T} (0) \equiv (1 \ \ 0 \ | \ 0 \ \ 0)
\ee
and (interchanging $a$ and $d$)

$$
\psi^{\rm T}(0) \equiv (0 \ \ 0 \ | \ 0 \ \ 1) \ .
\eqno {\rm(31')} $$
 
\h The spinorial quantities (31), (31'), together with the analogous ones
for $s_z = -{1 \over 2}$,
satisfy eq.(15): i.e., they are also solutions (in the CMF)
of the Dirac eigenvalue equation. \ This is the unique case in
which the zbw  disappears, while the
field spin is still present! In fact, even in terms of
eqs.(31)--(31'), one still gets that $\frac{1}{2} \ov{\psi} \Sig_z
\psi= +\frac{1}{2}$.\hfill\break
 
\h Since we have been discussing the classical limit ($v^\mu$) 
of a quantum quantity ($J^\mu$), let us add  that even the
well-known change in sign of the fermion wave function, under
360$^{\rm o}$-rotations around the $z$-axis, gets in our theory a 
natural classical interpretation. \ In fact, a 360$^{\rm o}$-rotation
of the coordinate frame around the $z$-axis (passive point of view) is indeed
equivalent to a 360$^{\rm o}$-rotation of the constituent $\cal Q$ around
the $z$-axis (active point of view). \ On the other hand, as a consequence of 
the latter transformation, the zbw angle $2 m\tau$ does vary of 360$^{\rm o}$,
 the
proper time $\tau$ does increase of a zbw period $T=\pi/m$,  and the
pointlike constituent does describe a complete circular orbit around
$z$-axis. At this point  it is straightforward to notice that, since 
the period $T= 2\pi / m$ of the $\psi(\tau)$ in
eqs.(14b)--(14c) is twice as large as the zbw orbital period, the wave
function of eqs.(14b)--(14c) does suffer a phase--variation of
180$^{\rm o}$ only, and then does change its sign: as it occurs in the standard
theory.
 
\h To conclude this Section, let us shortly consider the
interesting case obtained when {\em releasing} the conditions 
(22)--(25)
(and therefore abandoning the assumption of circular uniform motion), 
requiring instead that: 
\bb
|a|=|c| \quad\mbox{and}\quad |b|=|d| \ ,
\ee
so to obtain an internal oscillating motion along a constant straight line; 
where we understood  $\psi(0)$ to be written, as usual,
\[
\psi^{\rm T}(0) \equiv (a \ \ b \ | \ c \ \ d) \; .
\]
For instance, one may choose either
$$
\psi^{\rm T}(0) \equiv \frac{1}{\sqrt{2}} (1 \ \ 0 \ | \ 1 \ \ 0) \; ,
\eqno {\rm(32')} $$
 
or  \ $\psi^{\rm T}(0) \equiv \frac{1}{\sqrt{2}} (1 \ \ 0 \ | \ i \ \
0)$, \ or \ $\psi^{\rm T}(0) \equiv \frac{1}{2} (1 \ \ -1 \ | \ -1 \ \ 1)$,
 \ or \ $\psi^{\rm T}(0) \equiv \frac{1}{\sqrt{2}} (0 \ \ 1 \ | \ 0 \ \ 1)$,
 \ and so on.
 
\h In case (32'), for example, one actually gets
\[
< \vec{\ga} > \, = \, (0, 0, 1) \ ; \ < \vec{\al} > \, = \, (0, 0, 0)
\]
which, inserted into eqs.(24'a), (24'b), yield
\[
E^\mu =(0; 0, 0, 1) \; ; \qquad H^\mu = (0; 0,0,0) \; .
\]
Therefore, because of eq.(21a), we have now a linear,
oscillating motion [for which equations (22), (23), (24) and (25) do {\em
not} hold: here $V^2(\tau)$ does vary from 0 to 1!] along the $z$-axis:
\[
V_x(\tau)=0; \quad V_y(\tau)=0; \quad V_z(\tau)=\cos(2 m \tau) \ .
\]
 
\h This new case could describe an unpolarized, mixed state, since it is
\[
\sbf \equiv \frac{1}{2} \ov{\psi} \vec{\Sig}\psi = (0, 0, 0) \ ,
\]
in agreement with the existence of a linear oscillating motion.\\
 
{\bf 7. -- The light-like helical case.}\\
 
\h Let us go back to the case of circular uniform motions (in
the CMF), for which  conditions (22)-(25)  hold, with $\psi(0)$
given by eqs.(28). \ Let us fix our attention, however, on the special
case of light-like motion$^{[2]}$. \ The spinor fields $\psi(0)$,
corresponding to $V^2=0; \Vbf^2=1$, \  are given by eqs.(28) with \ 
$|a|=|d| \quad \mbox{for the}\quad s_z= +{1 \over 2} \quad\mbox{case}$, \
or \ $|b|=|c| \quad \mbox{for the}\quad s_z = -{1 \over 2} \quad\mbox{case}$; \
as it follows from eqs.(30) for $s_z = +{1 \over 2}$ and from the analogous
equations
$$
\Vbf^2=4|b^* c|=4|b|^2 (1-|b|^2)
\eqno {\rm (30'a)} $$
$$
\Abf^2=4 m^2 \Vbf^2 = 16 m^2 |b|^2 (1-|b|^2) \; ,
\eqno {\rm (30'b)} $$
 
holding for the case $s_z = -{1 \over 2}$. \ It can be easily shown that a 
difference in the phase
factors of $a$ and $d$ (or of $b$ and $c$, respectively) does not change 
the kinematics, nor the rotation direction, of the
motion;  but it does merely shift the zbw phase angle at $\tau = 0$. Thus, we
may choose {\em purely real} spinor components (as we did above). \  As
a consequence, the {\em simplest} spinors may be written as follows
$$
\psi_{(+)}^{\rm T} \equiv {\dis\frac{1}{\sqrt{2}}} (1 \ \ 0 \ | \ 0 \ \ 1)
\eqno {\rm (33a)} $$
$$
\psi_{(-)}^{\rm T} \equiv {\dis\frac{1}{\sqrt{2}}} (0 \ \ 1 \ | \ 1 \ \ 0)\
;
\eqno {\rm (33b)} $$
and then 
\[
< \vec{\ga} >_{(+)} = (1,0,0) \ ; \ < \vec{\al} >_{(+)} = (0,-i,0)
\]
\[
< \vec{\ga} >_{(-)} = (1,0,0) \ ; \ < \vec{\al} >_{(-)} = (0,i,0)     
\]
which, inserted into eqs.(24'), yield
\[
{E^\mu}_{(+)} = (0;1,0,0) \; ; \qquad{H^\mu}_{(+)} = (0;0,1,0)\;.
\]
\[
{E^\mu}_{(-)} = (0;1,0,0) \; ; \qquad{H^\mu}_{(-)} = (0;0,-1,0)\;.
\]
With regard to eqs.(33), let us observe that eq.(21a) implies 
for $s_z = +{1 \over 2}$ an {\em anti-clockwise\/}$^{[6,2]}$ internal motion, 
with respect to the chosen $z$-axis:
\setcounter{equation}{33}
\bb
V_x= \cos (2 m \tau); \quad V_y=\sin (2 m \tau); \quad V_z=0 \ ,
\ee
that is to say
 
$\hfill{\left\{ \dis \begin{array}{l}

X= {(2m)^{-1}} \sin (2 m\tau)+ X_0 \\

\vs{0.2 cm}

Y= - {(2m)^{-1}} \cos (2m \tau)+ Y_0 \\

Z=Z_0 \; ;
\end{array}\right.}
\hfill{\dis\begin{array}{r}
 \ \\ (34')\\ \ \end{array}}$
 
\

and a {\em clockwise} internal motion for $s_z = -{1 \over 2}$:

\bb
V_x=\cos (2m \tau);\quad V_y=-\sin(2 m\tau); \quad V_z=0 \; ,
\ee
that is to say
 
$\hfill{\left\{ \dis \begin{array}{l}

X= {(2m)^{-1}} \sin (2 m\tau)+ X_0 \\

\vs{0.2 cm}

Y= {(2m)^{-1}} \cos (2m \tau)+ Y_0 \\

Z=Z_0 \; .
\end{array}\right.}
\hfill{\dis\begin{array}{r}
\\ \\ (35') \\ \\ \end{array}}$
 
\h Let us explicitly observe that spinor (33a), related to
$s_z = +{1 \over 2}$ (i.e., to an anti-clockwise internal rotation), gets equal
weight contributions from the positive--frequency spin-up component
and from the negative--frequency spin-down component, in full
agreement with our ``reinterpretation" in terms of particles and
antiparticles, given in refs.[15]. \  Analogously, spinor (33b), related to
$s_z = -{1 \over 2}$ (i.e., to a clockwise internal rotation), gets
equal weight contributions from the positive-frequency spin-down
component and the negative-frequency spin-up component.
 
\h As we have seen above [cf. eq.(29)], in a {\em generic} reference frame
the polarized states are characterized by a helical uniform motion
around the $\impulse$-direction; thus, the $\la = +{1 \over 2} \;\;\; [\la =
-{1 \over 2}]$ spinor will
correspond  to an anti-clockwise [a clockwise] helical motion
with respect to the $\impulse$-direction.
 
\h Going back to the CMF, we have to remark that in this 
case eq.(24c) yields  for the zbw radius $R$ the traditional result:
\bb
R=\frac{|\Vbf|}{2 m} \equiv \frac{1}{2m} \equiv \frac{\la}{2} \; ,
\ee
where $\la$ is the Compton wave-length. \ Of course, $R = {1 / 2m}$
represents the maximum (CMF) size of the electron, among all the
uniform motion solutions ($A^2=$const; \ $V^2=$const);  the minimum,
$R=0$, corresponding to the Dirac case with no zbw ($V=A=0$), represented 
by eqs.(31), (31'): that is to say, the Dirac free
electron is a pointlike, extensionless object.
 
\h Finally, let us to underline that the present light-like solutions,
among all the uniform motion, polarized state solutions, are 
likely the most suitable solutions for a complete and ``realistic" (i.e., 
classically meaningful) picture of the free electron. \ Really,  the 
uniform circular
zbw motion with speed $c$ seems to be the sole that allows us ---if we think
the electric charge of the whole electron to be carried around by the internal
constituent $\cal Q$--- to obtain the electron Coulomb field and 
magnetic dipole (with the correct strength $\mu=e/2m$),
{\em simply by averaging} over a zbw period$^{[17,19}$ the electromagnetic 
field generated by the zbw circular motion of $\cal Q$ (and therefore 
oscillating with the zbw  frequency). \ Moreover, only in the light-like case
the electron spin can be actually regarded as totally arising from the internal
zbw motion, since the {\em intrinsic} term $\Delta^{\mu \nu}$ entering the
BZ theory$^{[1]}$ does {\em vanish} when $|\Vbf|$ tends to $c$.\\

{\bf 8. -- Generalization of the NDE for the non-free cases.}\\

\h Let us eventually pass to consider the presence of {\em external} 
electromagnetic
fields: $A^\mu \neq 0$. \ For the non-free case, Barut and Zanghi$^{[1]}$ 
proposed the following lagrangian

$$
\Le=\frac{1}{2} i (\dot{\ov{\psi}}\psi - \ov{\psi}\dot{\psi}) +
p_\mu(\dot{x}^\mu - \ov{\psi}\ga^\mu \psi) + eA_{\mu}\ov{\psi}\ga^{\mu}\psi
\eqno{\rm (37)}
$$
which in our opinion should be better rewritten in the following form, 
obtained directly from
the free lagrangian {\em via} the minimal prescription procedure:
$$
\Le=\frac{1}{2} i (\dot{\ov{\psi}}\psi - \ov{\psi}\dot{\psi}) +
(p_\mu - eA_{\mu})(\dot{x}^\mu - \ov{\psi}\ga^\mu \psi),
\eqno{\rm (38)}
$$
all quantities being expressed as functions of the (CMF) proper time $\tau$,
and the generalized impulse being now $p^\mu - eA^\mu$. \ We shall call as
usual $F^{\mu \nu} \equiv \pa^\mu A^\nu - \pa^\nu A^\mu$.\hfill\break
\h Lagrangian (38) does yield, in this case, the following system of differential equations:

\

$\hfill{\dis\left\{\begin{array}{l}
\dot{\psi}+ i (p_\mu - eA_\mu) \ga^\mu \psi=0\\

\dot{x}^\mu= \ov{\psi}\ga^\mu \psi\\

\dot{p}^\mu - e\dot{A}^\mu = e F^{\mu \nu} \dot{x}^\mu \; .
\end{array}\right.}
\hfill{\dis\begin{array}{r}
(39{\rm a}) \\ (39{\rm b}) \\ (39{\rm c}) \end{array}}$
 
\

As performed in Sect.{\bf 4}, we can insert the identity (10)
into eqs.(39a), (39b), and exploit the definition of the velocity field, 
eq.(39c). \
 We easily get the following {\em five} first--order {\em differential 
equations} (one scalar plus one vector equation) 
in the five independent variables $\psi$ and $p^\mu$:

\

$\hfill{\dis\left\{\begin{array}{l}
i(\ov{\psi}\ga^{\mu}\psi)\pa_{\mu}\psi = (p_\mu - eA_\mu) \ga^\mu \psi\\

(\ov{\psi}\ga^{\nu}\psi)\pa_\nu (p^\mu - eA^\mu) = e F^{\mu \nu} 
\ov{\psi} \ga_\nu \psi  \; ,
\end{array}\right.}
\hfill{\dis\begin{array}{r}
(40{\rm a}) \\ (40{\rm b}) \end{array}}$
 
\

which are now {\em field} equations (quantities $\psi,\ov{\psi},p$ and
$A$ being all functions of $x^\mu$).

\h The solutions $\psi(x)$ of system (40) may be now regarded as
the classical spinorial fields for relativistic spin-$\frac{1}{2}$ fermions,
in presence of an electromagnetic potential $A^\mu \neq 0$. 
We can obtain from eqs.(40) well-defined time evolutions, both for 
the CMF velocity $p^\mu / m $
and for the particle velocity $v^\mu$.  \ A priori, by imposing the
condition of finite motions, i.e., $v(\tau)$ and $p(\tau)$
periodic in time (and $\psi$ vanishing at spatial infinity), one will be 
able to find a discrete spectrum, out from the continuum  set of 
solutions of eqs.(40). \ Therefore, without solving any
eigenvalue equation, within our field theory we can individuate discrete 
spectra of energy levels for the stationary states, in analogy with what we
already found in the free case (in which the {\em uniform} motion condition
implied the z-componens $s_z$ of spin $s$ to be discrete). \ The case of a 
uniform, external magnetic field has been treated in ref.[16], where
---incidentally--- also the formal resolvent of system (39) is given. \ \ We
shall expand on this point elsewhere: having in mind, especially, 
the applications to classical problems,  
so as the hydrogen atom, the Stern and Gerlach experiment, the Zeeman effect, 
and tunnelling through a barrier.\\ 

{\bf 9. -- Conclusions.}\\

\h In this paper we have analyzed and studied the electron 
internal motions  (predicted by the BZ theory), as functions of the 
initial conditions, for both time-like and space-like speeds.
\ In so doing, we revealed the noticeable {\em new} kinematical properties 
of the velocity field for the spinning electron.  For example, we found that 
quantity $v^2 \equiv v_{\mu}v^{\mu}$ may assume any real value and be variable
in time, differently from the ordinary (scalar particle)
case in which it is always either $v^2 = 1$ or $v^2 = 0$. \ By requiring
$v^2$ to be constant in time (uniform motions), we found however that 
[since the CMF does {\em not} coincide with the rest frame!] quantity $v^2$ 
can assume all values in the interval 0 to 1, 
depending on the value of $\psi(0)$.\\ 
\h Moreover, this paper shows clearly the correlation existing between 
electron polarization and Zitterbewegung. \ In particular, 
in Sect.{\bf 6} we show that the requirement
${\bf s} = (0,0,\pm {1\over 2})$ \ 
[i.e., that the classical spin magnitude, corresponding 
to the average quantum spin magnitude, be ${1\over 2}$] \ corresponds to 
the requirement of uniform motions for $\cal Q$, and vice-versa. 
We found the zbw oscillation to be a uniform circular motion, with frequency
$\om = 2m$, and with an orbital radius 
$R = |\vbf|/2m$  that in the case of light-like orbits equals half the
Compton wave-length; \ the clockwise (anti-clockwise) helical 
motions  corresponding to the spin-up (spin-down) case.\\
\h We introduced also an equation for the 4-velocity field (different 
from the corresponding equation of Dirac theory), which 
allows an intelligible description of the electron internal motions:
thus overcoming the well-known problems about the 
physical meaning of the
Dirac {\em position operator} and its time evolution.$^[20]$  Indeed, the 
equation for the standard Dirac case
was devoid of any
intuitive kinematical meaning, because of the appearance of the imaginary 
unit $i$ in front of the acceleration (which is related to the 
non-hermiticity of the velocity operator in that theory).\\
\h Finally, we have considered a natural generalization of the NDE for 
the non-free
case, which allows a classical description of the interaction between 
a relativistic
fermion and an external electromagnetic field (a description we shall 
deepen elsewhere).\\

\

\

{\bf Acknowledgements}\\

The authors acknowledge continuous stimulating discussions with P. Cardieri, P.L. Dias Peres, W. de Freitas Filho, A. Pagano, 
M. Pav\v{s}i\v{c}, M. Zamboni-Rached, and particularly Hugo E. Hern/'andez-Figueroa. \  Thanks for helpful collaboration and interest 
are also due to C. Giardini, I. Licata, P. Riva, C. Rizzi, A. Sakaji and particularly 
S. Paleari. At last, we would like to recall the work
performed by Waldyr A. Rodrigues Jr. (recently passed away), who contributed to the creation of research groups in theoretical physics, which paid particular attention to the structure of spinning particles, to magnetic poles, and to the use of Clifford algebras in space-time physics. \

\

\end{document}